\documentclass[12pt]{iopart}
\usepackage{graphicx}
\usepackage{subcaption}
\usepackage{tikz}
\usetikzlibrary{external}
\usetikzlibrary{matrix,shapes,arrows,positioning,chains,calc}

\begin{document}

\title[]{Mimicking complex dislocation dynamics by interaction networks}

\author{Henri Salmenjoki$^1$, Mikko J. Alava$^1$, and Lasse Laurson$^{1,2}$}

\address{$^1$Aalto University, Department of Applied Physics, PO Box 11000, 00076 Aalto, Finland}
\address{$^2$Laboratory of Physics, Tampere University of Technology, P.O. Box 692, FI-33101 Tampere, Finland}
\ead{henri.salmenjoki@aalto.fi}
\begin{abstract}
Two-dimensional discrete dislocation models exhibit complex dynamics in relaxation and under external loading. This is manifested both in the time-dependent velocities of individual dislocations and in the ensemble response, the strain rate. Here we study how well this complexity may be reproduced using so-called Interaction Networks, an Artificial Intelligence method for learning the dynamics of complex interacting systems. We test how to learn such networks using creep data, and show results on reproducing individual and collective dislocation velocities. The quality of reproducing the interaction kernel is discussed.

\end{abstract}

\maketitle

\section{Introduction}

Lately the use of Artificial Intelligence and Machine Learning (ML) in understanding complex physics problems has received quite some attention \cite{zdeborova-2017}. There are many reasons like that reducing large quantities of data may allow to detect new features from parameterized potentials for atomistics \cite{behler-2016,botu-2015,huan-2017}, or to find descriptors or labels for dynamics that otherwise pose challenges \cite{wiewel-2018,pathak-2018,koch-2018}, or that one may classify physical systems to phases like in statistical physics and learn about the transitions and order parameters \cite{carrasquilla-2017,vanniewenburg-2017,wetzel-2017}. Very recent work has also applied ML methods in the realm of plastic deformation \cite{papanikolaou-2018}. Here, we exploit the idea of Interaction Networks, that have gained interest as a means of learning and reproducing the dynamics of complex, interacting systems \cite{battaglia-2016}. Our interest here is the question what do we learn by applying such a tool to interacting dislocation systems, in particular for the paradigmatic case of two-dimensional discrete dislocation models \cite{miguel-2001,zaiser-2006,laurson-2010,ispanovity-2014,janicevic-2015,papanikolaou-2017}. As a reminder, these are collections of interacting particles with a topological charge (the Burgers vector) and a very complicated long-range two-particle interaction force.

The idea of the IN model is first to learn how physical objects of the system interact, and then to learn the relation between the interactions and the dynamics of the objects. 
Thus, the IN provides two applicable features for physics research: First, it can learn the underlying interactions and this way extract a physical model from the  system. Second, it can predict the development of the system with the acquired dynamics.
In the case of the dislocation system of the simulations, the dynamics simply reduce to the velocities of the dislocations. Learning the interaction from quasistatically driven simulations proved to be more challenging, so the IN was trained with data from simulations with constant external stress instead. In the rest of this paper, we first outline the methodology and network learning, and then present the results and a short summary.

\section{Methods}

\subsection{Two-dimensional Discrete Dislocation Dynamics (DDD) model}

As an interacting dislocation system, we consider a 2D DDD model similar to the one 
in Refs.~\cite{miguel-2001,zaiser-2006,laurson-2010,ispanovity-2014},
describing a set of parallel, straight edge dislocations with 
equal number of positive and negative Burgers vectors of magnitude $b$. 
The dislocations move in a square box of size $L$ with periodic boundaries 
and interact via the dislocation-generated shear stress fields \cite{hirth-1982},
\begin{equation}
\sigma_d(\textbf{r}) = \sigma_d(x,y) = D b \frac{x(x^2-y^2)}{(x^2+y^2)^2},
\label{eq:dislocation_stress}
\end{equation}
where $D=\mu/2\pi(1-\nu)$, with $\mu$ and $\nu$ the shear modulus and Poisson
ratio, respectively. 
Taking into account the infinite amount of periodic images, the exact form of the dislocation interactions is \cite{hirth-1982}
\begin{equation}
\sigma_d (x,y) = \frac{\pi D b}{L} \frac{\sin \left(\frac{2 \pi x}{L} \right) \left[\cosh \left(\frac{2 \pi y}{L} \right) - \cos \left(\frac{2 \pi x}{L} \right) - \frac{2 \pi y}{L} \sinh \left(\frac{2 \pi y}{L} \right) \right]}{\left[ \cosh \left(\frac{2 \pi y}{L} \right)-\cos \left( \frac{2 \pi x}{L} \right)\right]^2}.
\label{eq:infim_stress}
\end{equation}
The overdamped equations of motion describing the glide
motion of dislocations along the $x$ direction are
\begin{equation}
\frac{v_i}{\chi b}  = s_i b [ \sigma_{\mathrm{ext}} + \sum_{i \neq j} s_j \sigma_d (\textbf{r}_j-\textbf{r}_i) ],
\label{eq:dislocation_velocity}
\end{equation}
where $\chi$ is the dislocation mobility, $s_i$ and $s_j$ are the signs of the 
Burgers vectors of dislocations $i$ and $j$, respectively, and 
$\sigma_{\mathrm{ext}}$ is the external stress. We measure lengths in 
units of $b$, time in units of $1/\chi b D$ and stresses in units of $D$. 
To mimic dislocation annihilation, two dislocations with opposite Burgers 
vectors are removed from the system if their distance is less than $b$.

\subsection{Interaction network}

For predicting the dynamics the IN implementation follows directly the network presented in  Ref.~\cite{battaglia-2016}. The dynamics to be learned and reproduced is the velocities of the dislocations. Initially, we tested quasistatic simulations with a stress ramp, but these turned out to be harder than the case of ordinary creep. In quasistatic loading, the dislocations experience two extreme states of total jamming during the stress ramp and rapid flow during the strain bursts. The dislocations are jammed most of the simulation time, so to obtain proper training set with adequate amount of states with dislocation motion would have required a lot of simulations and clever sampling of the data. Therefore, the IN was trained with data from simulations with constant external stress instead \cite{miguel-2001,laurson-2010}. 

The input of the IN consists of a full description of the system which in this case means the state variables of the dislocation positions $x$ and $y$, Burgers vector sign $s$ and a variable indicating whether the dislocation is annihilated. These are collected to matrix $\textbf{O}$ which is of size $D_S \times N_O$ where $D_S$ is the number of state variables (here this is four) and $N_O$ is the number of dislocations.  
The state matrix is fed to IN with matrix $\textbf{X}$ representing external effects and two permutation matrices, $\textbf{R}_R$ and  $\textbf{R}_S$. 
$\textbf{X}$ is a $1 \times N_O$ matrix, comprising of the external stress acting on each dislocation, while the permutation matrices are both $N_O \times N_R$, where $N_R$ denotes the number of relations. As all the dislocations are interacting with each other, $N_R = N_O (N_O-1)$. The subscripts $R$ and $S$ refer to receiver and sender of the interaction, respectively. $\textbf{R}_R$ and  $\textbf{R}_S$ are constructed so that every dislocation pair is taken into account exactly once. 

\begin{figure}[h]
\centering
\includegraphics[width=0.8\textwidth]{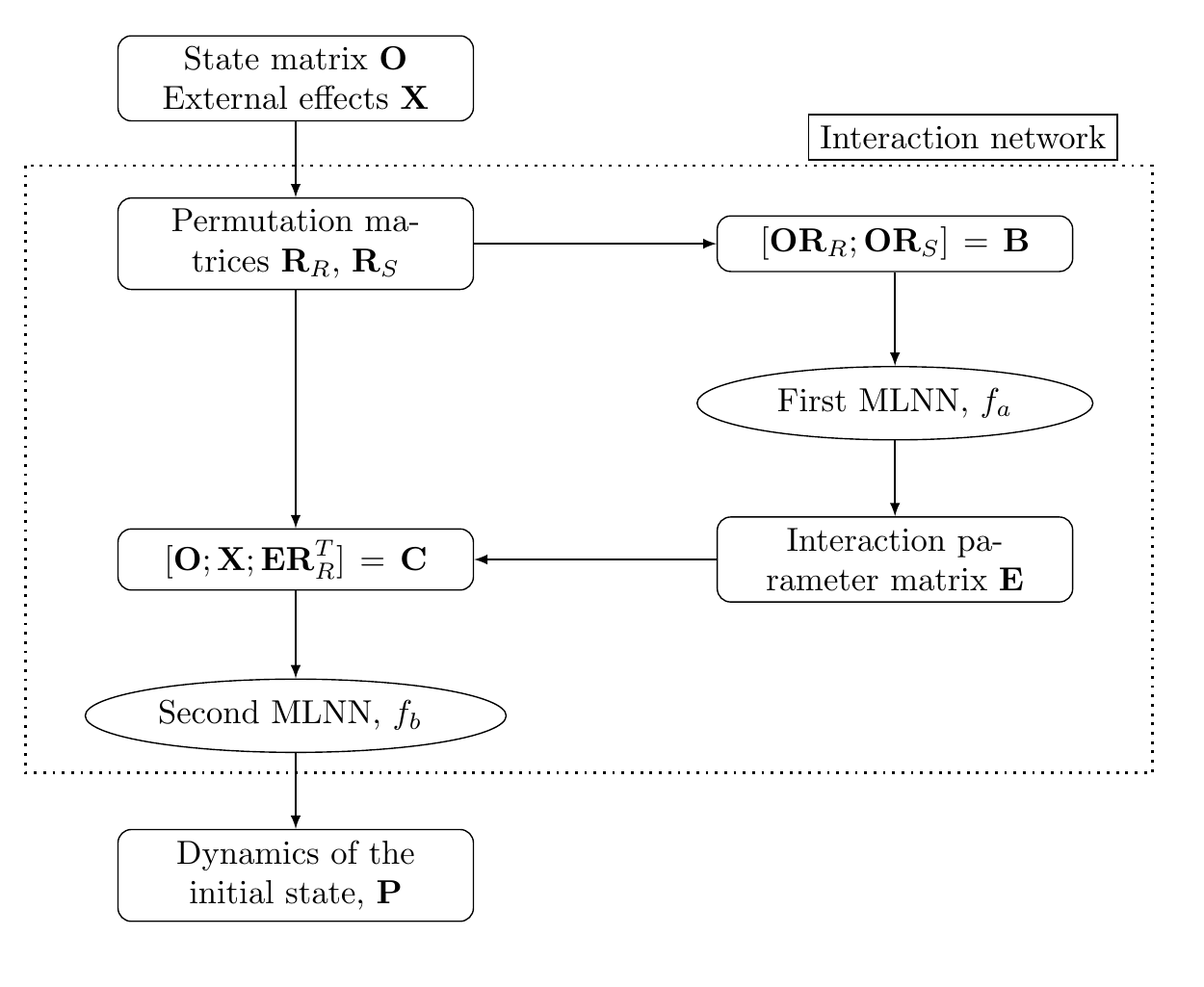}
\caption{Flow chart of the IN. \label{fig:flowchart} }
\end{figure}

Step-by-step guide of the IN operating principles is presented in Figure. \ref{fig:flowchart}
The first step in IN is to calculate matrix multiplications  $\textbf{OR}_R$ and  $\textbf{OR}_S$, as these represent the state of every receiver and sender of the interactions. Then, these are concatenated to form a $2D_S \times N_R$ matrix $\textbf{B}$ (considering the interaction term, it is evident that the sign of the receiver is irrelevant and erasing column of the receiving dislocation sign from $\textbf{B}$ slightly improves IN performance). $\textbf{B}$ acts as input to the first multi-layered neural network (MLNN) of the IN, $f_a$. The output from $f_a$ is a matrix $\textbf{E}$ of size $D_E \times N_R$, where the  $N_R$ interactions are individually represented with $D_E$ parameters. Afterwards, $\textbf{E}$ is multiplied with $\textbf{R}_R^T$ to form $D_E \times N_O$ matrix $ \bar{\textbf{E}}$  which contains the total interactions  acting on a dislocation.
Finally, matrices $\textbf{O}$, $\textbf{X}$ and $\bar{\textbf{E}}$ are concatenated to a $(Ds+1+D_E) \times N_O$ matrix $\textbf{C}$ (again for the purposes of this work, $\textbf{O}$ includes inessential information concerning the velocity, as the absolute position of the dislocation does not affect the velocity and, thus, this is discarded). This is fed into the second MLNN, $f_b$, which outputs the vector $\mathbf{P}$ representing the velocity of every dislocation. 

\subsection{Training the network}

Training the IN proved to be substantially complex. The approach of training with systems of few objects and then scaling upwards, which was taken in \cite{battaglia-2016}, was here impossible due to the poor variance in structures of few dislocations. In the small systems, the dislocations avoided configurations of close neighbors. 
Moreover, scaling the same IN to larger systems was practically impossible, considering that the simulated dislocation interactions are dependent on the system size $L$ due to the periodic boundaries as seen in Eq.\ \ref{eq:infim_stress} \cite{hirth-1982}. Hence, the training was conducted with systems initialized with 100 dislocations in a box with $L=50b$. For the training set, 225 initialized dislocation systems were simulated with constant stresses ranging from $0.11$ to $0.17$, and from these simulations, 50000 images were chosen randomly. The test set was gathered from 25 simulations with the corresponding stresses. 
Due to the large number of dislocations per image, the training was time-consuming. Therefore, proper optimization of the MLNNs $f_a$ and $f_b$ was omitted as only a few architectures were tested.   
For the final implementation, both MLNNs consisted of three hidden layers, first with 100 and second with 50 neurons. Additionally, the number of parameters for interaction representation was set to $D_E=10$. In what follows, the predictions and their quality are measured with the score typical for this kind of network testing defined as
\begin{equation}
S= 1 - \frac{\sum_i (d_i - y_i)^2}{\sum_i (d_i - \langle d_i \rangle)^2}
\end{equation}
where $d_i$ is the (desired) output and $y_i$ is the IN output for a given sample $i$. Obviously, the score $S$ ranges from 1 (perfect prediction) to 0 (use average response) to minus infinity.

\section{Results}
\subsection{Predicting individual dislocation velocities with the interaction network \label{section:predictions_in}}

\begin{figure}[h!]
\centering
\includegraphics[width=0.9\textwidth]{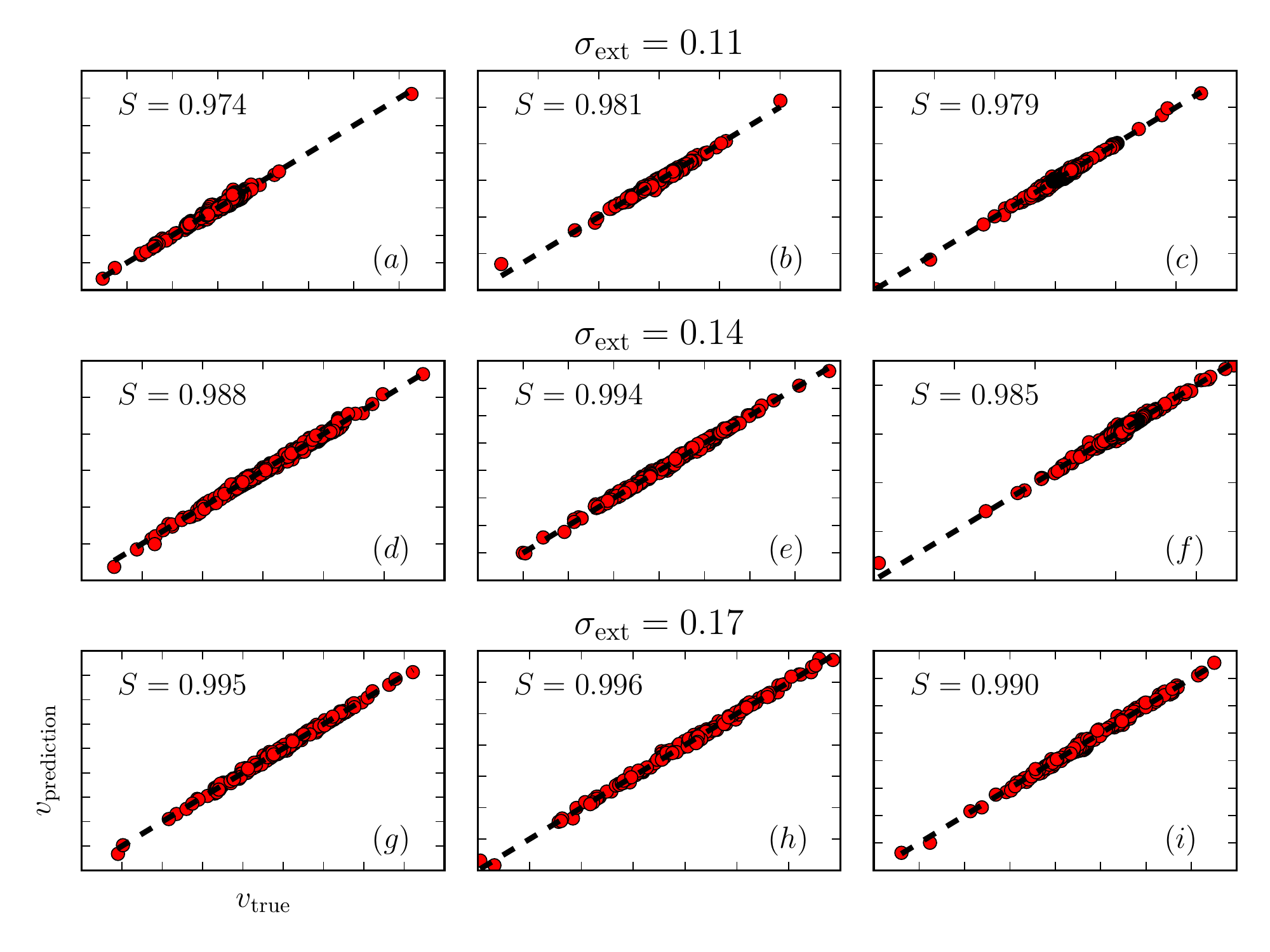}
\caption{ IN predictions plotted as a function of the actual velocities of the single dislocations in test systems $(a)-(i)$. For systems $(a)-(c)$,  $(d)-(f)$ and $(g)-(i)$ the applied external stress was set to   $\sigma_{\mathrm{ext}}=0.11$, $\sigma_{\mathrm{ext}}=0.14$ and $\sigma_{\mathrm{ext}}=0.17$, respectively. The dashed lines represent $v_{\mathrm{prediction}}=v_{\mathrm{true}}$.\label{fig:in_predictions_separate}}
\end{figure}

IN managed to learn the dislocation interactions quite well.  
Figure \ref{fig:in_predictions_separate} shows the predictions versus the actual single dislocation velocities in nine test systems, $(a)-(i)$. The velocities were calculated for a random set of images from the simulation and the images were treated as separate cases. Now, the score $S$ settled near unity as the actual values coincided well with the network output. Additionally, the score appeared to be higher with larger $\sigma_{\mathrm{ext}}$.

\subsection{Integrating dislocation systems with IN}
\begin{figure}[h!]
\centering
\includegraphics[width=0.9\textwidth]{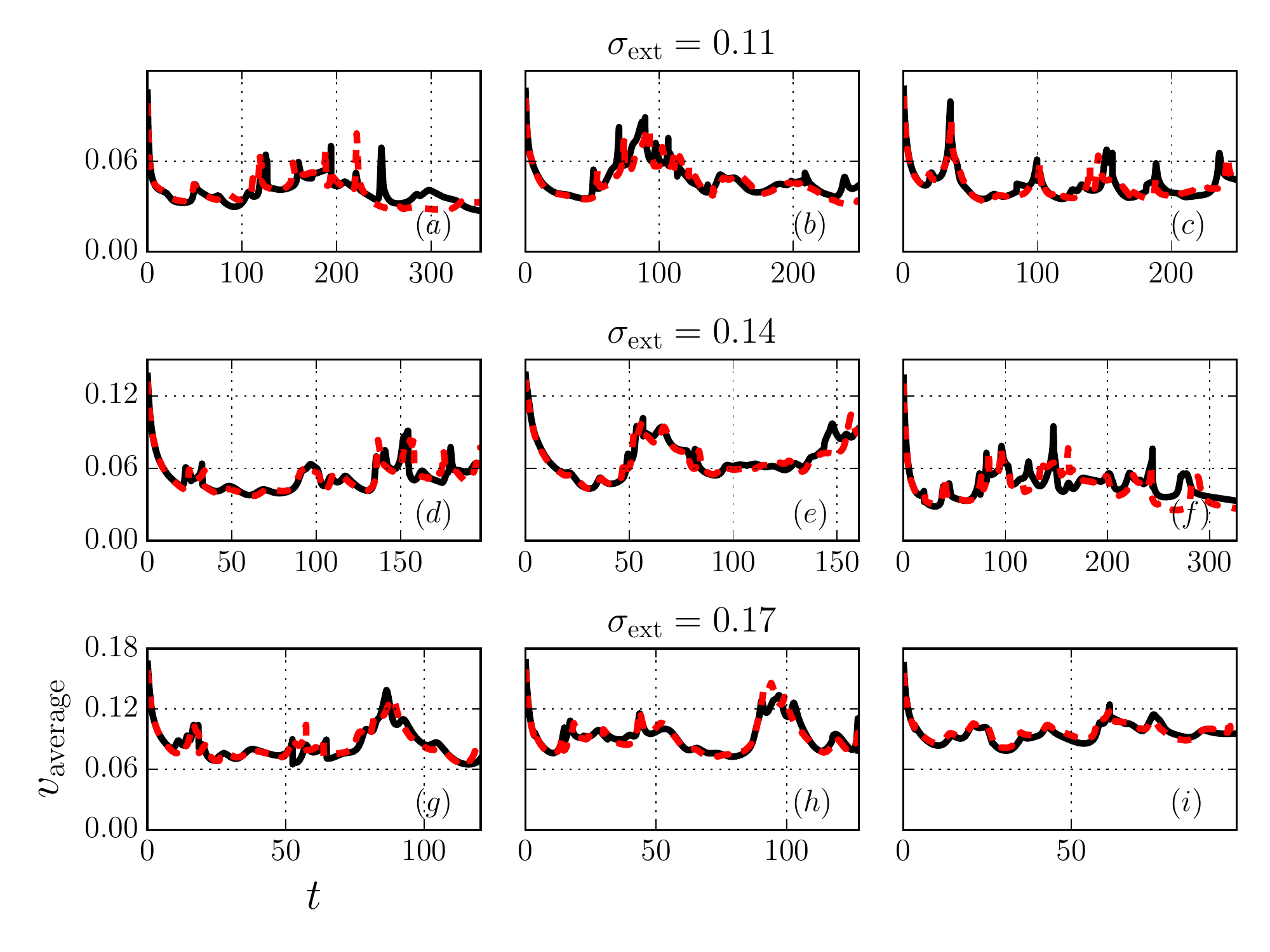}
\caption{ $v_{\mathrm{average}}$ plotted as a function of time $t$ from the dynamic predictions (dashed red line) produced by the IN for the test systems $(a)-(i)$    compared with the simulation results (solid black line). Simulations were computed until the strain in system reached $\varepsilon=0.2$. \label{fig:in_predictions_dynamic}}
\end{figure}

Starting from the initial dislocation positions, IN could also generate dynamic predictions. In practice, this was easy to implement with a time-integration routine that updated the dislocations with the velocities from IN and revised the possible dislocation annihilations after every step. In Figure \ref{fig:in_predictions_dynamic}, the average velocity $v_{\mathrm{average}}$ of the dislocations is plotted for the test systems $(a)-(i)$ from the simulation and the dynamic prediction. Although $v_{\mathrm{average}}$ is not an absolute measure of how well the predicted and true dynamics coincide, it is used here as a measuring stick instead of comparing for instance individual dislocation positions or velocities at every time step as issues might ensue due to different outcomes for annihilation. 
The figure shows that generally the predictions coincide with the simulations, especially in the start of the loading. Most predictions seem to capture the simulation dynamics as the velocity curve shapes are similar, but slightly shifted. In these systems, the dislocations indeed reach similar final configurations when compared separately. But in some systems, for instance $(a)$ and $(f)$, the predictions fail towards the end of the loading.  Naturally, this is due to the small errors in the velocity predictions, that accumulate to push the system into a totally different configuration.

\subsection{Interaction kernels learned}
\begin{figure}[h]
    \centering
    \begin{subfigure}[b]{0.49\textwidth}
        \includegraphics[width=\textwidth,trim={1cm 0 1cm 0},clip]{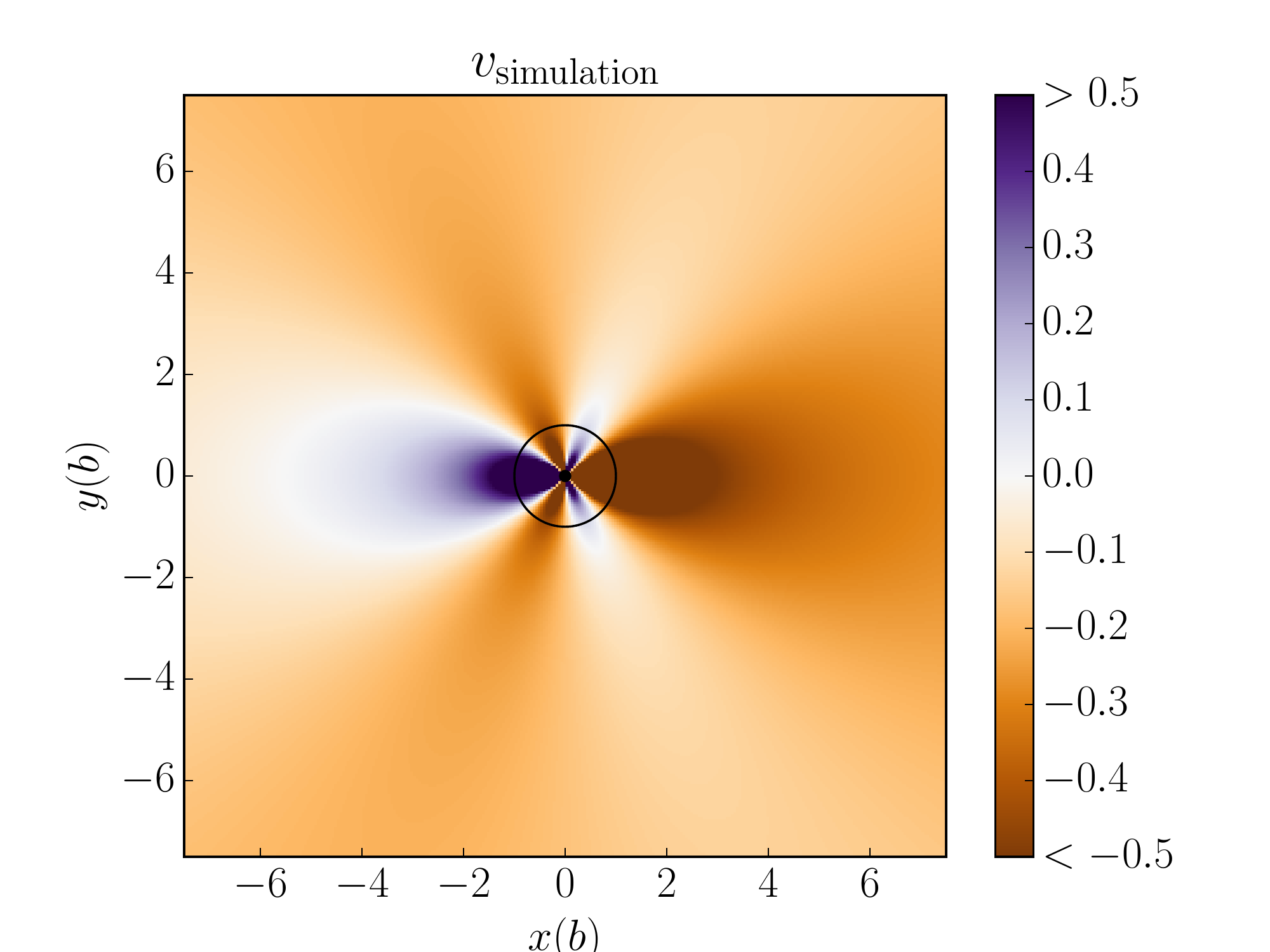}
        \label{fig:sd_sf_sim_diff}
        \caption{ }
    \end{subfigure}
    \begin{subfigure}[b]{0.49\textwidth}
        \includegraphics[width=\textwidth,trim={1cm 0 1cm 0},clip]{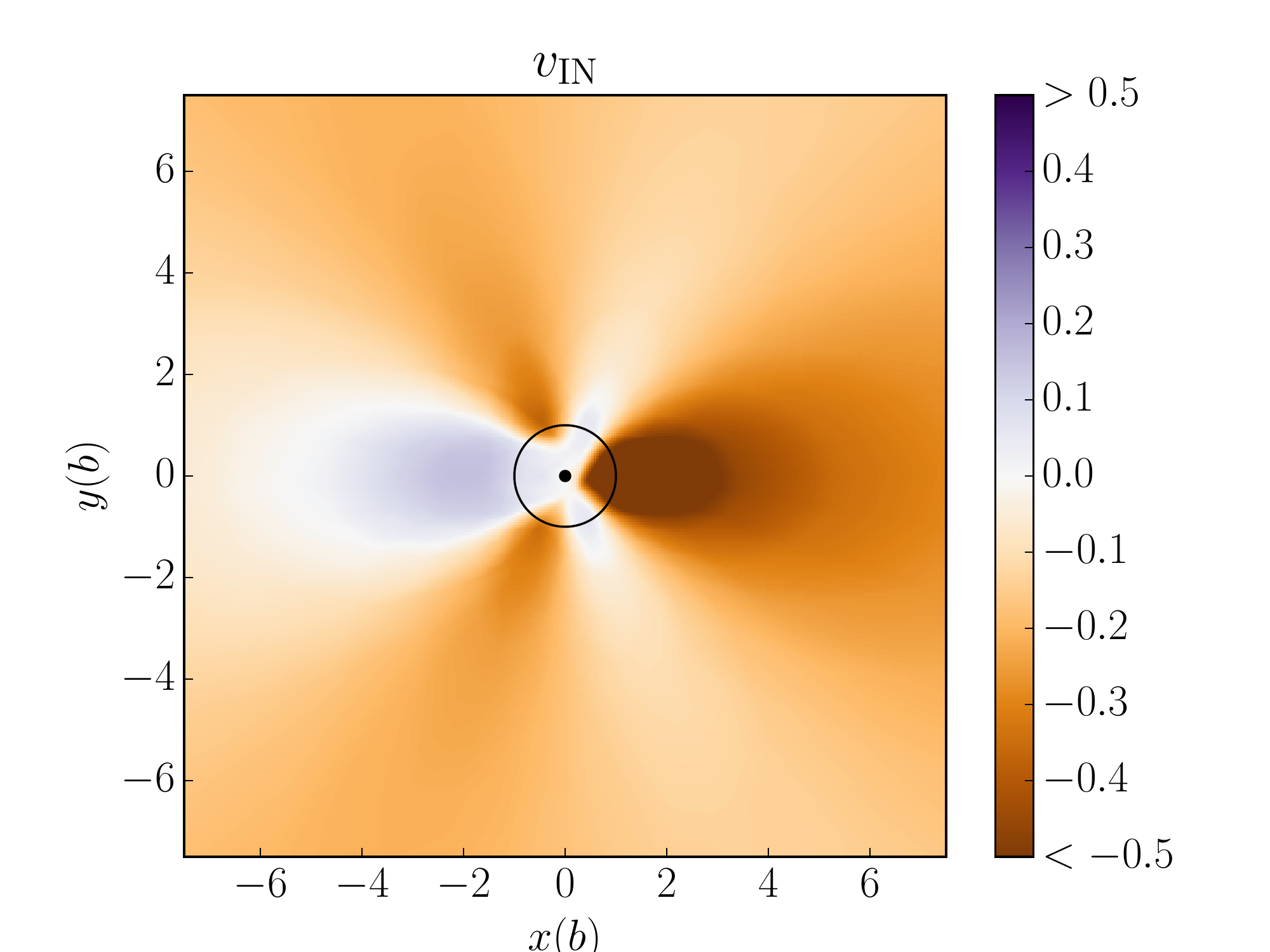}
        \label{fig:sd_sf_in_diff}
        \caption{ }
    \end{subfigure}
    \begin{subfigure}[b]{0.49\textwidth}
        \includegraphics[width=\textwidth,trim={1cm 0 1cm 0},clip]{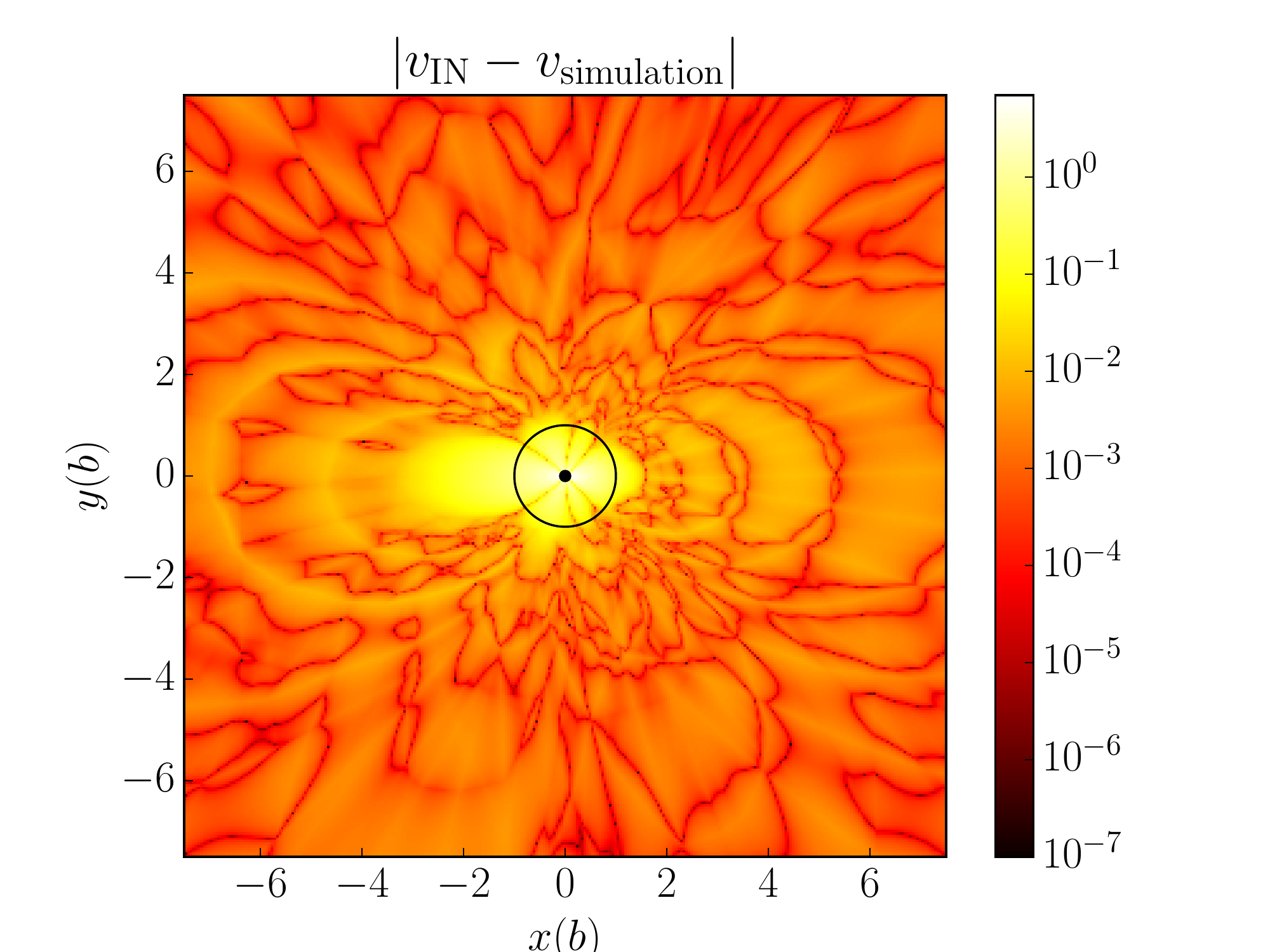}
        \label{fig:sd_sf_abse_diff}
        \caption{ }
    \end{subfigure}    
    \begin{subfigure}[b]{0.49\textwidth}
        \includegraphics[width=\textwidth,trim={1cm 0 1cm 0},clip]{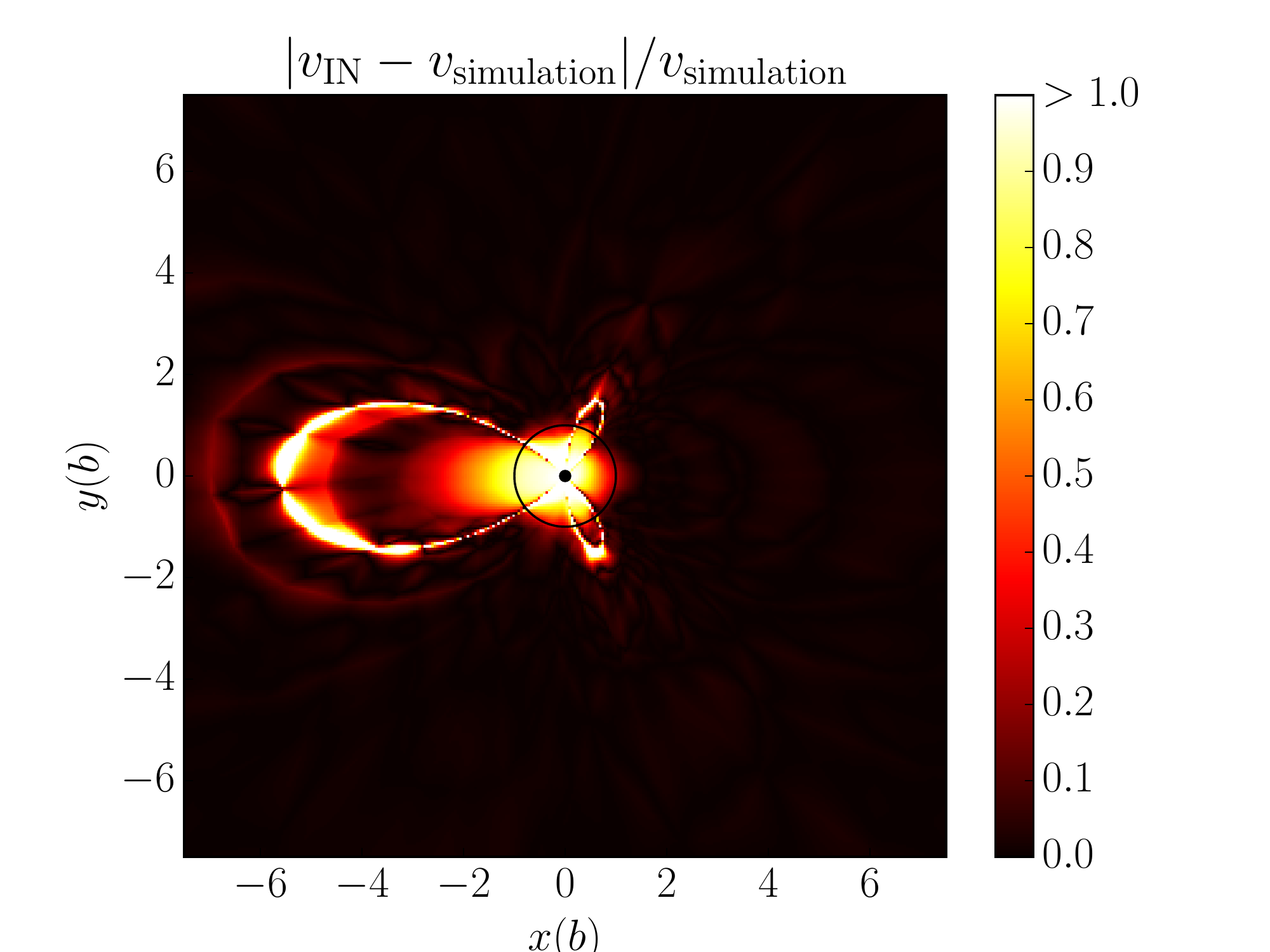}
        \label{fig:sd_sf_rele_diff}
        \caption{ }
    \end{subfigure}
    
    \caption{(a) Exact velocity field of Equation (\ref{eq:dislocation_velocity}), (b) IN prediction of the field, (c) absolute difference and (d) relative difference between the fields. The test dislocation, marked with the black dot, was assigned with a positive Burgers vector, while the field is for a dislocation with negative Burgers vector. The black circle shows the core of the test dislocation, where the dislocations would annihilate. External stress was set to $\sigma_{\mathrm{ext}}=0.17$.}\label{fig:sd_sf_diff}
\end{figure}

To obtain an overview of the interaction IN learned, velocity field output of IN was computed near a test dislocation and compared with the exact field of Equation \ref{eq:dislocation_velocity}. 
The test dislocation was assigned with a positive Burgers vector and it was placed to the origin. Figure \ref{fig:sd_sf_diff} illustrates the exact and IN velocity fields along with the absolute and relative error between the fields for a dislocation with negative Burgers vector. 
A glance at the fields suggests that the IN has learned the interaction particularly well, when the dislocations are separated by some distance, while the learned field near the test dislocation is not as strong as the exact field. This is confirmed by the error fields. A simple explanation would be that the training set does not include many images with oppositely signed dislocations extremely close to each other on almost the same glide plane. This is due to the fact that the approaching dislocations experience the largest velocities that lead them to annihilate.

Interestingly, the absolute error is largest left from the test dislocation, while the field above, right and below the test dislocation has been acquired by the IN.   
Indeed, closer inspection of the error figures indicates that the largest errors occur in the regions where the velocity is reversed to the external field, that is where the negative dislocation travels to the positive $x$-direction.
On the other side of the test dislocation, the IN has had no such problems.
Again, this is most likely arising from the training set, where dislocations traveling to the 'wrong' direction form a minority. Thus for the IN, learning  the field on the other side of the test dislocation seems to be easier as the dislocation velocities are there to the 'assumed' direction.

\begin{figure}[h]
    \centering
    \begin{subfigure}[b]{0.49\textwidth}
        \includegraphics[width=\textwidth,trim={1cm 0 1cm 0},clip]{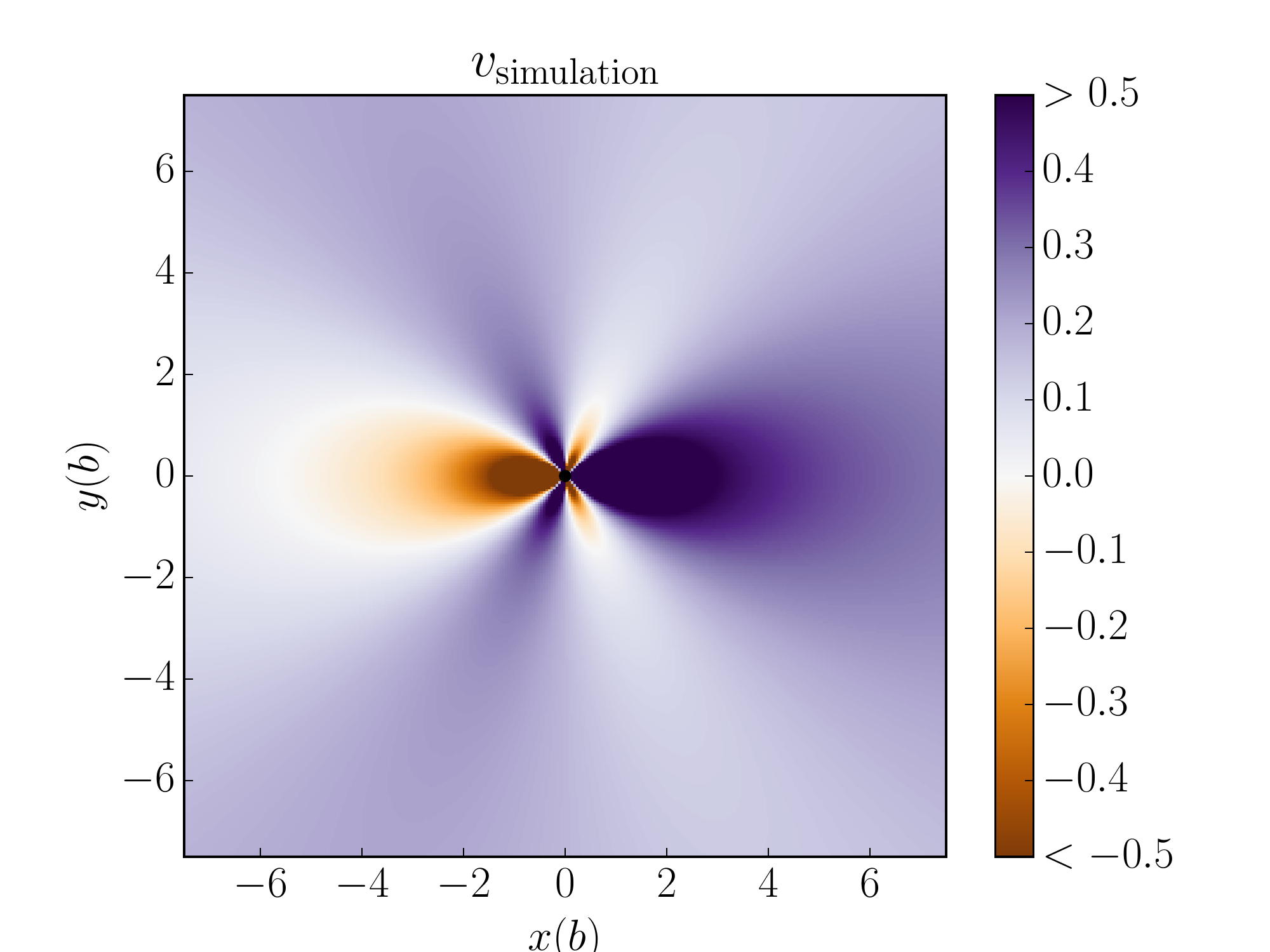}
        \label{fig:sd_sf_sim_same}
        \caption{ }
    \end{subfigure}
    \begin{subfigure}[b]{0.49\textwidth}
        \includegraphics[width=\textwidth,trim={1cm 0 1cm 0},clip]{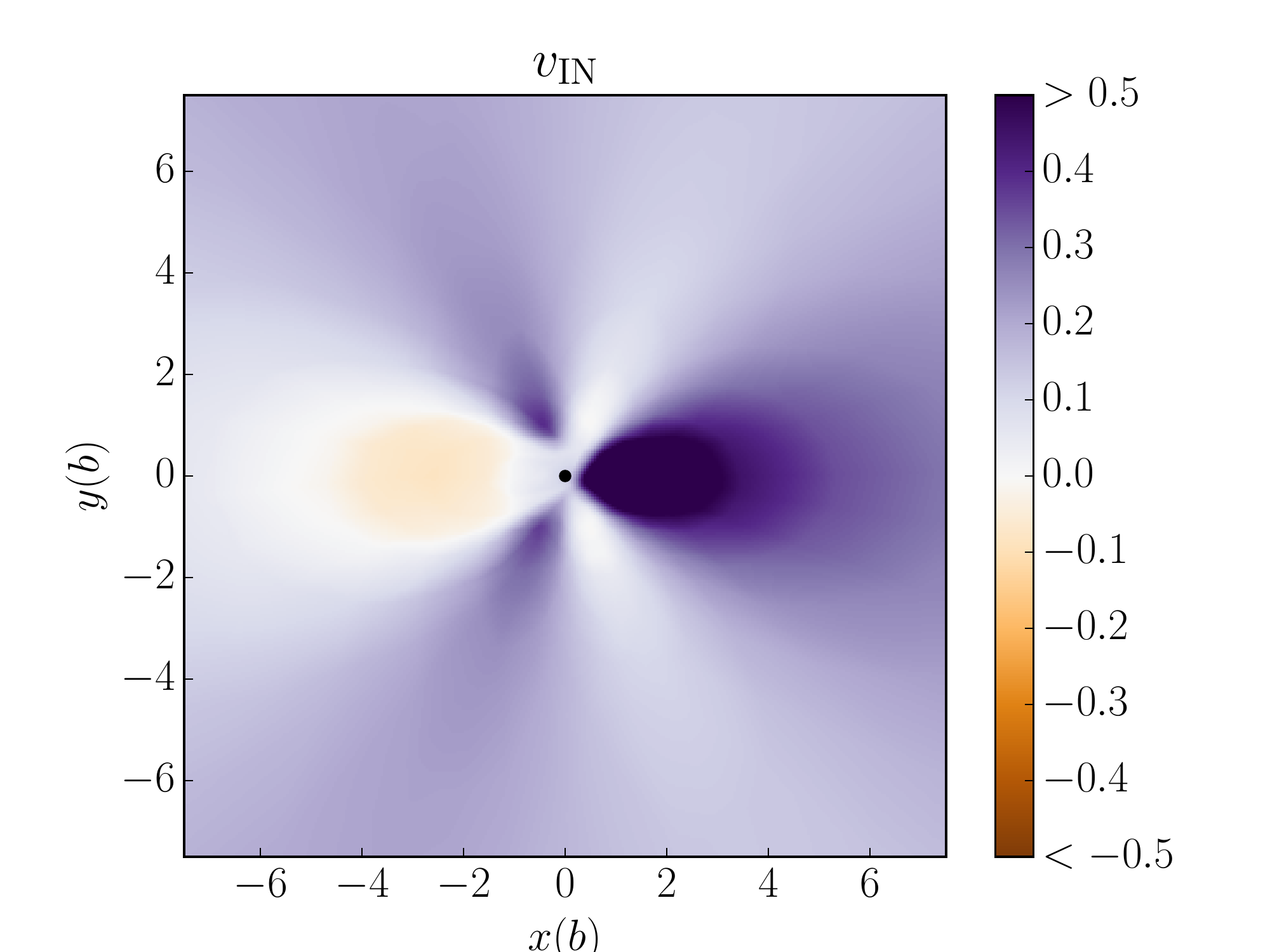}
        \label{fig:sd_sf_in_same}
        \caption{ }
    \end{subfigure}
    \begin{subfigure}[b]{0.49\textwidth}
        \includegraphics[width=\textwidth,trim={1cm 0 1cm 0},clip]{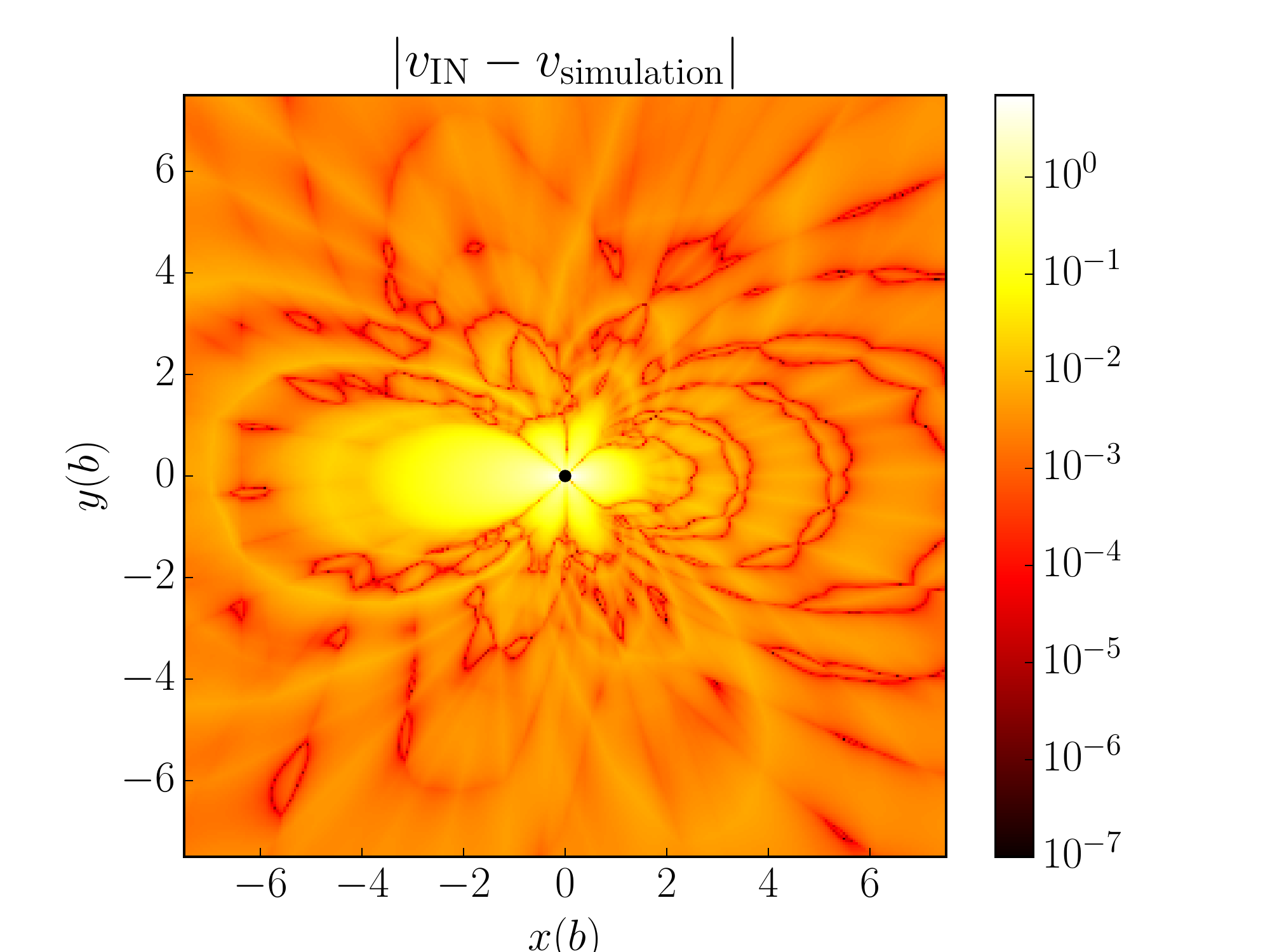}
        \label{fig:sd_sf_abse_same}
        \caption{ }
    \end{subfigure}    
    \begin{subfigure}[b]{0.49\textwidth}
        \includegraphics[width=\textwidth,trim={1cm 0 1cm 0},clip]{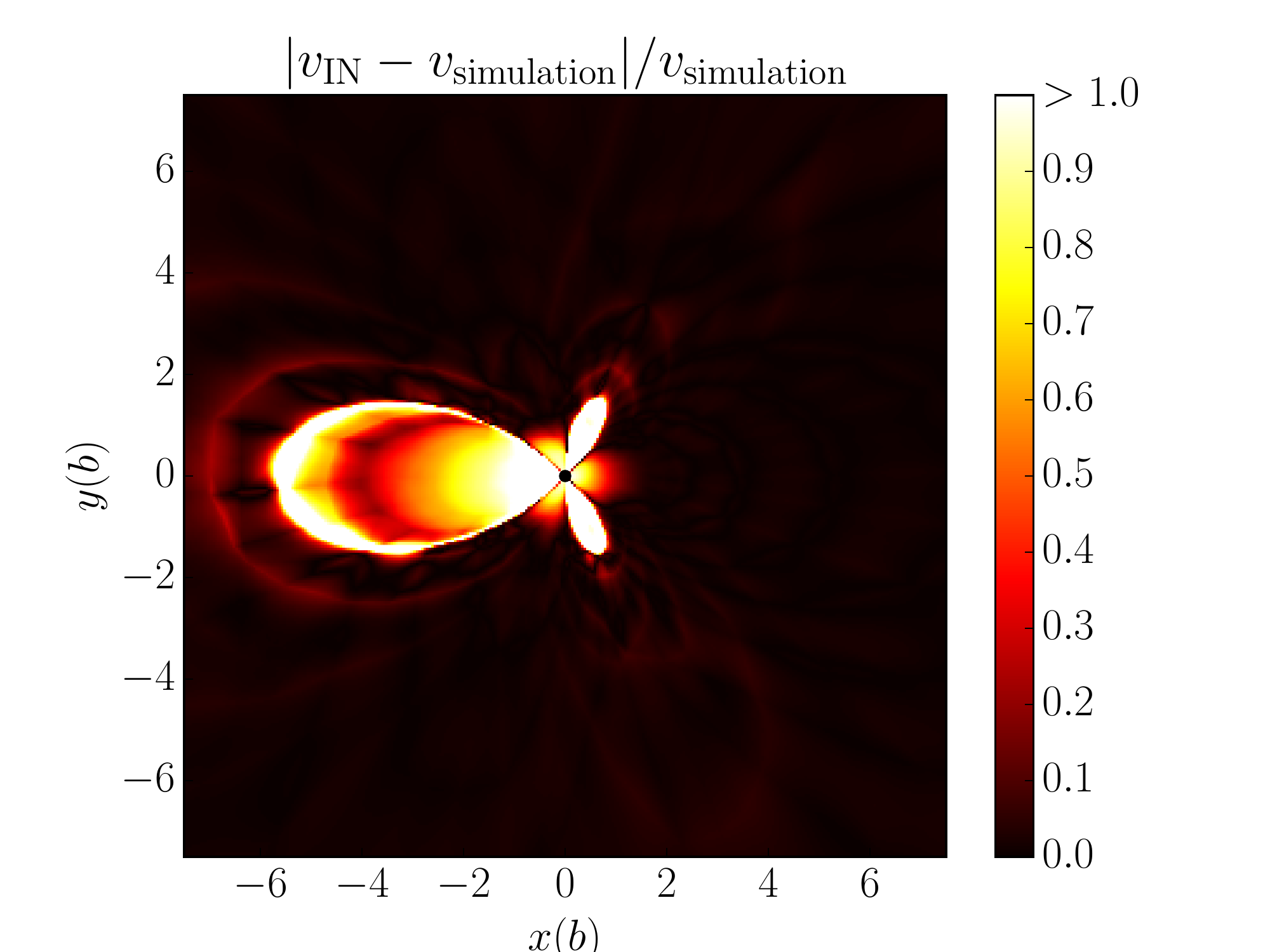}
        \label{fig:sd_sf_rele_same}
        \caption{ }
    \end{subfigure}
    
    \caption{(a) Exact velocity field of Equation (\ref{eq:dislocation_velocity}), (b) IN prediction of the field, (c) absolute difference and (d) relative difference between the fields. Now, both of the dislocations have positive Burgers vector and $\sigma_{\mathrm{ext}}=0.17$. }\label{fig:sd_sf_same}
\end{figure}

Correspondingly, Figure \ref{fig:sd_sf_same} shows the velocity fields and their difference for dislocations with positive Burgers vector. The relative error is again more significant for regions where the velocity should be against the external field.  The effect is emphasized with the same signed dislocations, as they strongly repel each other and seldom form close configurations. Therefore, in the case where the dislocations reach these rare configurations during creep, the IN starts to falter and change from the actual simulation. 
In addition to larger training set and better network architecture,  another possible way to improve the IN performance could be to use even larger dislocation systems, because there the possible dislocation configurations are more diverse.
\section{Summary}
Dislocation dynamics, even in the simplest case of two dimensions, presents a complex system  due to the frustration and long-range, anisotropic interaction forces present. We thus wanted to mimick or apply Artificial Intelligence -based learning strategies for dynamical systems to yielding.

Our study could in principle be generalized to other dimensions. In 3D crystals, this is made somewhat more complicated by the fact that the IN should learn not only a scalar stress field as in our case, but the full stress tensor. Moreover, under multi-slip conditions the mapping from stress field to dislocation velocity needs to take into account the different resolved shear stresses on different glide planes. On the other hand, 1D models (of dislocation pile-ups \cite{moretti}) could constitute a more straightforward avenue of future work as there the stress field to be learned is a 1D scalar function (instead of the 2D function in the present case), although in that case one might want to consider the model in the presence of a quenched pinning force landscape (to make the problem more interesting, as is done in Ref. \cite{moretti}) which presumably would create some complications for the IN. 

It turns out that in spite of the deceptive complexity - avalanches, intermittency all well-known in the collective dislocation behaviour - Interaction Networks are able to reproduce both individual dislocations' response and the summed total creep rate or average velocity. The way this works as usual improves with more and more learning data. The reason for this is in a way probably elementary: even though the O$(N^2)$ -like dynamics is complicated, very high dimensional neural networks present an embedding which has enough freedom for an accurate reproduction. There are two obvious (to us) questions. What would a detailed study of avalanches (as reproduced by IN, or the difference in their description) lead to? Could one eventually study the complexity of 2D DDD models by simply teaching well this kind of models, that are then easier numerically to run, despite the fact that evaluating the trained IN and traditional simulations scale similarly with system size \cite{wiewel-2018,pathak-2018}? \\

\noindent
{\bf Acknowledgements.} This work has been supported by the Academy of Finland through 
an Academy Research Fellowship (LL, project no. 268302) and the Centers of Excellence program
(HS, project no. 251748). We acknowledge the computational resources provided by the Aalto 
University School of Science “Science-IT” project, as well as those 
provided by CSC (Finland).\\

\noindent
{\bf Author contributions.} 
H.S., M.J.A. and L.L. designed the study. H.S. performed the numerical simulations and implemented the interaction network. H.S. wrote the first draft of the manuscript, and all authors contributed to the final form of the manuscript.\\

\noindent
{\bf Data availability.} The data that support the findings of this study are 
available from the corresponding authors on reasonable request.

\end{document}